# A Picture's Worth a Thousand Words: Visualizing n-Dimensional Overlap in Logistic Regression Models with Empirical Likelihood

Paul A. Roediger (*proediger@comcast.net*)

**Abstract**. In this note, conditions for the existence and uniqueness of the maximum likelihood estimate for multidimensional predictor, binary response models are introduced from a sensitivity testing point of view. The well-known condition of Silvapulle is translated to be an empirical likelihood maximization which, with existing R code, mechanizes the process of assessing overlap status. The translation shifts the meaning of overlap, defined by geometrical properties of the two-predictor groups, from the intersection of their convex cones is non-empty to the more understandable requirement that the convex hull of their differences contains zero. The code is applied to reveal the character of overlap by examining minimal overlapping structures and cataloging them in dimensions fewer than four. Rules to generate minimal higher dimensional structures which account for overlap are provided. Supplementary materials are available online.

**Comment**. Contains 1 pdf file consisting of 22 pages, 13 tables, 10 figures, and 1 appendix.

## 1. Introduction

In a typical sensitivity test, experimenters sequentially stress specimens of a system or product, and observe a binary response (0 or 1). The stresses are called predictors. Regardless of response, the test is considered destructive, so survivors are never tested twice. It's a last resort kind of test (binary responses are least informative), but for many commodities, it's the only one available, so test efficiency, e.g., as measured by achieving overlap in the minimum average number of runs, is essential. Predictors are almost always univariate because, to date, there are no procedures to efficiently vary multiple factors in order to achieve overlap. From a research perspective, progress is "very slow on 2D" (Wu, personal communication, April 17, 2018).

 The initial goals of a sensitivity testing strategy are to achieve mixed results and ultimately produce two overlapping predictor groups, $X^1$ (Case) and $X^0$ (Non-Case). The achievement of both goals is a necessity for the maximum likelihood estimate (MLE) to exist and be unique. The Neyer Test (Neyer 1994) and the 3pod procedure (Wu and Tian 2014) are the two most popular strategies to efficiently achieve overlap. They have been implemented in *gonogo.R* (available for download at https://www2.isye.gatech.edu/~jeffwu/sensitivitytesting/), documented (Roediger 2018), applied (Ray and Roediger 2018) and compared (Wang, Tian and Wu 2020). Once overlapping data is attained, models are fitted sequentially to refine the parameter estimates, which also allows the next stress to be computed via, e.g., a D-optimality criterion.

There are many applications for this type of testing, including, but not limited to: characterizing the sensitivity of explosives; determining safe or effective drug dosages; evaluating chemical toxicity; and ruggedizing commercial products and packaging. Other applications are discussed in Ray and Roediger (2018).

## 2. Empirical Likelihood (EL) for Overlap – Background

Sensitivity testing is just one of the many application areas where binary responses are the rule. In a broader context it has been said that "characterizing the existence and uniqueness of the MLE in logistic regression has been a classical problem in statistics" (Candès and Sur 2020).

The data considered in this paper consists of $n$ binomial responses, $y_i$, observed at predictor rows $x_i \in R^d$. Responses are linked to predictors via $\Pr[y_i = 1 | x_i] = G(\alpha + \beta \cdot x_i)$ where $\alpha \in R$ (the intercept term) and $\beta \in R^d$ (a row vector of slope coefficients). The model $G$ is typically a cumulative distribution function. Table 1 lists the necessary and sufficient conditions (Silvapulle 1981) for the log likelihood function $l(\alpha, \beta) = \sum_{i=1}^{n} \left( y_i \log(G) + (1 - y_i) \log(1 - G) \right)$ to have a unique maximum likelihood estimate. Two conditions target the model and two target the data.

**Table 1**. Silvapulle's Conditions for the Existence and Uniqueness of the MLE

| Target | Condition | Detail |
|---|---|---|
| Model | Monotonicity | $G$ is strictly increasing at every value of $z$ with $0 < G(z) < 1$ |
| Model | Convexity | $\log(G(z))$ and $-\log(1 - G(z))$ are convex functions of $z$ |
| Data | Full Rank | data is the extended predictor matrix $\mathrm{E} = (e_i)_{n \times (d+1)}$, where $e_i = (1, x_i)$ |
| Data | Overlap | $S \cap F = \phi$ where $S = \left\{ \sum_{i: y_i = 1} u_{1i} x_{1i} \mid u_{1i} > 0 \right\}$, $F = \left\{ \sum_{i: y_i = 1} u_{0i} x_{0i} \mid u_{0i} > 0 \right\}$ $x_1 = x_{y=1}$, $x_0 = x_{y=0}$ and $\sum_{i: y_i = 1} u_{1i} = \sum_{i: y_0 = 0} u_{0i} = 1$ |

A more useable overlap condition, one that doesn't rely on an intersecting convex cone concept, was proposed by Albert and Anderson (1984). It promised that linear programming (LP) techniques could demonstrate overlap by showing no hyperplane separates the two extended predictor groups. They went on to distinguish two kinds of separation, quasi-complete and complete. The former results when the data has mixed responses, is neither overlapping nor completely separated, and possesses at least one Case and one Non-Case observation on the hyperplane. Interestingly, Zeng (2017) offers a rigorous proof that the two interpretations are equivalent, even though this technical matter was covered by others, including Rossi (1984). Widespread use of a reliable LP implementation was made possible much later in the R package *safeBinaryRegression* (Konis 2007).

A better tool is developed in the present work, one which depends upon Empirical Likelihood (EL). It was inspired in part by re-reading Rossi (1984), who gave the following LP conditions for separated data in 1D: $b_0 + b_1 x_{i|y_i=1} > 0$ and $b_0 + b_1 x_{i|y_i=0} < 0$. This meant of course that the set of differences $\delta = x_{i|y_i=1} - x_{i|y_i=0}$, a complimentary Minkowski sum, couldn't straddle zero. In 2017, after experimenting with higher dimensional differences, reading Davis [2014], and noticing a minor error in our paper (Owen and Roediger 2014) , the present author wrote to Art Owen about this, and mentioned the following new idea: "If the non-zero, normalized predictor differences, Cases minus Non-Cases, could be constrained to lie on the interior of a unit hemisphere, then the data separates completely. Otherwise it does not".  Art responded: "This is the 'empirical likelihood' problem, and I have essentially bullet-proof  R-code for it" (Owen, personal communication, March 20, 2017).  The "new idea" was apparently not so new. In fact, a decade earlier, he summed up Silvapulle's criterion this way:

> "The cone intersections may seem unnatural. A ***more readily interpretable condition*** (emphasis added) is that the relative interior of the convex hull of the x's for y = 0 intersects that for y = 1" (Owen 2007).

The quantity $\delta$ has been extensively studied in the 3D algorithmic literature pertaining to robotics and computer-aided-design (Gilbert, Johnson and Keerthi 1988) and collision detection and game physics (van den Bergen 2004).

In the next section an EL overlap test based upon the R function *emplik* (Owen, 2013) is described. This sophisticated code is posted at http://statweb.stanford.edu/~owen/empirical/ as *scel.R*. Unlike *safeBinaryRegression*, a new test (*elstat*) assesses overlap status and distinguishes two types of separation, complete and quasi-complete. Because EL optimizations applied to assessing overlap status are so closely related to the Silvapulle criteria, the new approach has considerable pedagogical value.

*Emplik* also comes in a modified form, *cemplik*, to take on counts. Documentation (*countnotes.pdf*) and the *cemplik* code (*scelcount.R*) are also available on Owen's website. The counts are treated as weights and don't have to be natural numbers (Owen 2017). This version is applicable wherever weighted logistic regressions are widely used, e.g., credit scoring with reject inference (Zeng 2017).

The standard reference on empirical likelihood is Owen (2001). There, anyone unfamiliar with the subject will profit by discovering the many interesting applications of these methods. It is suspected that the message of EL for overlap isn't well-known. It would be worthwhile to add overlap assessment to the list of important EL applications.

### 3.   The EL Data-related Conditions for Existence and Uniqueness of an MLE

Let *X*  be a  *n*  by  *d*  matrix. Each row of *X* is a vector of predictor values for one observation.
The rank of a set of vectors will be understood to be the rank of a matrix containing those rows as vectors. Let $X^1$ and $X^0$ be the  $n_1$ by *d*  and  $n_0$ by *d*  matrices of predictor rows having Case and Non-Case responses, respectively. We assume throughout that  $n_1 n_0 > 0$ , i.e., the responses are not all the same. Let $\delta$  be the matrix of  $n_1 n_0$  predictor displacements (as rows)  $X_i^1 - X_j^0$ . The EL data-related conditions are given in Table 2.

**Table 2**. EL Conditions Comparable to Silvapulle's Two Data-Related Conditions

| Condition | Detail |
|---|---|
| Full Rank | the rank of $\delta$ is *d* |
| Overlap | the origin lies in the interior of the convex hull of $\delta$ |

**Theorem 1**. The data-related conditions of Table 1 and Table 2 are equivalent

**Proof**. Consider the EL problem of determining whether the origin lies within the interior of the convex hull of $\delta$. When true (Owen 2001, 2013), there are unique weights, $w_{ij}$, which maximize $\prod_{i=1}^{n_1}\prod_{j=1}^{n_0} n_1 n_0 w_{ij}$, subject to $\sum_{i=1}^{n_1}\sum_{j=1}^{n_0} w_{ij}\delta_{ij} = 0$, $w_{ij} > 0$, and $\sum_{i=1}^{n_1}\sum_{j=1}^{n_0} w_{ij} = 1$. Then, the two points $S = u_1 X^1$ and $F = u_0 X^0$ are identical, where $u_{1i} = \sum_{j=1}^{n_0} w_{ij} > 0$ and $u_{0j} = \sum_{i=1}^{n_1} w_{ij} > 0$. Silvapulle's overlap condition holds because $\sum_{i=1}^{n_1} u_{1i} = \sum_{j=1}^{n_0} u_{0j} = 1$. Since rows of $\delta$ are row differences of $\mathrm{E} = (1, x)_{n \times d+1}$ (per Table 1), it follows that $rank(\mathrm{E}) - 1 = rank(\delta)$ and the full ranks of $\mathrm{E}$ and $\delta$ each occur together. In the reverse direction, if the Silvapulle condition is satisfied, the $u_{1i}$'s and $u_{0j}$'s provide weights $w_{ij} = u_{1i} u_{0j}$ solving the EL optimization problem.

## 4. EL tools

R scripts to demonstrate EL overlap, along with documentation, are included as supplementary materials (*elSup.R* and *el4overlapSup.pdf*). In this EL-for-Overlap regime, sensitivity data is always stored as a list object having three components: an n by d matrix of predictors (as rows) named "x"; a vector of responses named "y"; and a vector of row ID's named "rid". With *L* as such a list, the function *elstat*(L) returns overlap status as determined by *emplik0* ($\delta$), where $\delta$, the matrix of predictor displacements, is returned by the function *deex*(L). Table 3 lists important functions pertinent to assessing overlap status and cataloging minimal overlap configurations.

**Table 3**. Important EL Functions

| Function | Provides | Reference | Function | Provides | Reference |
|---|---|---|---|---|---|
| *deex*(L) | Displacements $\delta$ | Theorem 1 | intrm(L) | Interim Form $V$ | Table 8 |
| emplik0($\delta$) | Weights $W$ | Theorem 1 | *stdf*($d_1, d_0$) | Standard Form $\Lambda$ | Table 8 |
| elstat(L) | Overlap Status | Tables 4,5 | equid($d_1, d_0$) | Equidistant Form $E$ | Section 7 |
| *eldef*(L) | Deflated $L$ | Sections 5,6 | *tictactoe*() | Type II Search | Section 13 |
| *tdat1*(i) | Examples | Figs. 2,3,7,8 | *mixw*(L) | Randomize $L$ | Section 14 |
| *config3*(L) | Configuration ID | Figs. 2,3,7,8 | *rquaz*(n) | Quasi of size n | Section 14 |
| *esimp*(d) | Equi-simplex $\Delta^{(d)}$ | Figure 1 | *safetest*(L) | Overlap Pass/Fail | Section 14 |

## 5. An Example

The version of *emplik* used in this work is *emplik0*, a slight modification of the original (without prints). *emplik0* is called from other functions, such as *elstat,* to determine overlap status; and *eldef,* to produce minimal subsets according to overlap status. Table 4 presents a sample data set accompanied with output from *elstat* which demonstrates that the $S \cap F \neq \phi$ criterion is indeed satisfied.

**Table 4**. EL Overlap Demonstration of $X = W_0$ using *elstat*

| $u_1^T$ | $X^1$ | | | $u_0^T$ | $X^0$ | | |
|---|---|---|---|---|---|---|---|
| .16493310 | 9.0 | 5.0 | 5.0 | .07232476 | 6.7 | 5.8 | 4.0 |
| .08925699 | 10.0 | 4.4 | 9.0 | .12967985 | 9.5 | 4.7 | 7.0 |
| .21147676 | 8.4 | 4.2 | 3.5 | .13076825 | 10.1 | 2.6 | 3.0 |
| .08701106 | 11.9 | 4.1 | 6.0 | .08068072 | 7.5 | 4.2 | 2.5 |
| .08695881 | 11.2 | 4.7 | 8.0 | .11694525 | 7.3 | 5.1 | 2.1 |
| .09646835 | 9.0 | 4.8 | 10.0 | .07262221 | 10.4 | 4.1 | 3.5 |
| .08728623 | 11.0 | 4.6 | 8.1 | .08199628 | 8.4 | 4.6 | 2.0 |
| .17660869 | 7.0 | 2.6 | 4.1 | .31498268 | 10.0 | 4.0 | 11.0 |
| $S = u_1 X^1$ | 9.227367 | 4.194798 | 5.981682 | $F = u_0 X^0$ | 9.227367 | 4.194798 | 5.981682 |

**Note**: The data set $W_0$ is provided in the supplementary materials

The call to *elstat* amounts to performing an empirical likelihood test that the true mean of $\delta$ is zero as described in Owen (2001). The actual mean of $\delta$ is $\bar{x}_{y=1} - \bar{x}_{y=0}$, the same quantity that determines the sign of the slope coefficient in a logistic regression (Owen and Roediger, 2014).

In practice, overlap status depends on the rank of $\delta$ and total weight $w_{tot} = \sum_{i=1}^{n_1} \sum_{j=1}^{n_0} w_{ij}$, and is assessed in accordance with Table 5.

**Table 5.** Status of Data as a Function of $d$ and $w_{tot}$

| rank($\delta$) | $w_{tot} < \varepsilon$ | $\varepsilon \leq w_{tot} \leq 1-\varepsilon$ | $w_{tot} > 1-\varepsilon$ |
|---|---|---|---|
| $d$ | Complete Separation | Quasi Separation | Overlap |
| $< d$ | Complete Separation | Quasi Separation | Quasi Separation |

**Note**: Use, e.g., $\varepsilon = 10^{-8}$

## 6. Minimal Overlap Configurations

In this note *emplik* is used to sequentially drop runs from known overlapping data sets until overlap is destroyed. An abundance of curiosity was the motivating factor, to get a better picture of what constitutes overlap in dimension two, the next frontier for modern sensitivity testing strategies. Regarding this run-dropping, *emplik0* resolves a problem described by Demidenko (2001): "from a robustness point of view, this yields a problem. Many robust estimators are constructed such that outlying points are deleted or appropriately down-weighted. However, it can happen that the whole data set has overlap but the reduced data set does not."

Minimal overlap configurations are found to have many interesting properties. It was demonstrated in Owen and Roediger (2014, Lemma 2) that Silvapulle's full rank and overlap conditions imply that $n \geq d+1$. This result is extended to include an upper limit in the following:

**Theorem 2**. Let $n$ be the number of predictors in a minimal overlap configuration. Then $d+2 \leq n \leq 2(d+1)$.

**Proof**. The lower limit is addressed in Owen and Roediger (2014). For the upper limit, let $1 \leq n_2' \leq min(n_1, n_0)$ be the number of unique doubletons belonging to $X^1 \cap X^0$. This leaves $n_1'$ and $n_0'$ unduplicated singletons belonging to $X^1$ and $X^0$, respectively. To satisfy Silvapulle's full rank condition, $n_{unique} = n_1' + n_0' + n_2' \geq d+1$. To satisfy Silvapulle's overlap condition, $n = n_0' + n_1' + 2n_2' \geq d+2$. Since $n_2'$ can vary from 0 and $d+1$, it follows $n$ is limited to values between $d+2$ and $2(d+1)$ inclusive.

**Corollary 1**. There is only one minimal overlap configuration which maximizes $n$ in $2(d+1)$ runs: it consists of $d+1$ doubletons at design points possessing full rank.

**Definition.** Minimal overlap configurations of dimension $d$ that achieve overlap in $n = d+2$ runs are termed Type I configurations and are denoted by $C_I^d$. Minimal overlap configurations consisting of $n > d+2$ runs are termed Type II configurations and are denoted by $C_{II}^d$. When type is unspecified, $C_{I/II}^d$ suffices to indicate a minimal overlap configuration.

We see in the next sections that the two types are structurally different: in regard to the presence of lower dimensional Type I and/or Type II sub-configurations, Type I configurations contain none, whereas every run in a Type II configuration belongs to at least one.

The R function *eldef* is used to isolate subsets of overlapping data that cannot be further reduced. With it, numerical examples of the two types of configurations are identified as subsets of $W_0$ depicted in Table 4. First, the last two rows on the left and the last three rows on the right provide an example of a Type I configuration. It meets the overlap condition because a line segment connecting two points in the Case data set intersects the interior of a triangle made of three points in the Non-Case data set (see configuration *f* in Figure 2). A Type II configuration is also found in the first three rows on the left and the first three rows on the right. Overlap in this case is assured because one point in the Non-Case data set is in the interior of a line segment connecting two points in the Case data set; one point in the Case data set is in the interior of a line segment connecting two points in the Non-Case data set; and the two line segments are non-coplanar (see configuration $A_9$ in Figure 7). These two subsets are defined (as $W_1$ and $W_2$) in the supplementary materials. Other examples of Type II configuration are the multiple doubletons mentioned in Corollary 1.

Type I and Type II configurations are size-constrained in accordance with Table 6.

**Table 6**. Sizes of minimally-overlapping configurations

| $d$ | $size(C_I^d)$ | $size(C_{II}^d)$ | | | |
|---|---|---|---|---|---|
| 0 | 2* | | | | |
| 1 | 3 | 4* | | | |
| 2 | 4 | 5 | 6* | | |
| 3 | 5 | 6 | 7 | 8* | |
| 4 | 6 | 7 | 8 | 9 | 10* |

**Note**: Starred entries consist of $d+1$ doubletons

The small run sizes involved make minimal overlapping configurations manageable objects to explore with our EL tools.

## 7. Type I Configurations

The appropriate model for a Type I configuration consists of two overlapping simplexes, one for the Case data and one for the Non-Case data. Recall: A simplex, usually denoted by $\Delta^{(d)}$ or $\Delta^d$, is defined to be the convex hull of $d+1$ affinely-independent vertices $\{P_i \in R^d\}_{i=0}^{d}$. Affine independence (and non-degeneracy) means the $d$ differences $v_i \in \{P_i - P_0\}_{i=1}^{d}$ are linearly independent and the d-dimensional volume of the simplex is non-zero. For Type I configurations the overlapping simplex model for $X^0$ and $X^1$ means $rank(X_0) = n_0 - 1$, $rank(X_1) = n_1 - 1$ and $rank(X) = d = n - 2$.

The smallest simplexes are depicted in Figure 1.

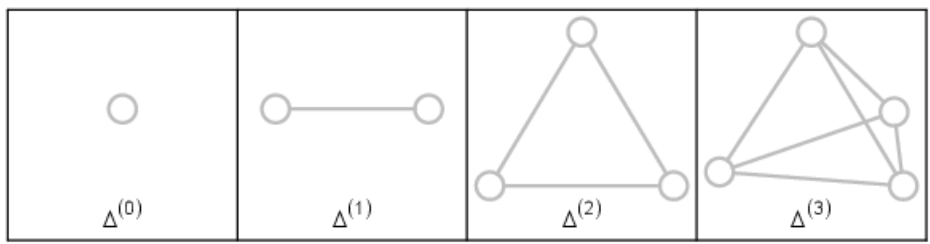

**Figure 1**: The first four simplexes

In Figure 1, the center of each circle represents a design point. For graphical purposes, each circle is ready to receive one of three icons: "X" to indicate a Case (or 1) response; "O" to indicate a Non-Case (or 0) response; or "$\otimes$", to indicate a Case response and a Non-Case response (i.e., a doubleton). In terms of the icons, a catalog of Type I configurations in dimensions fewer than four is presented in Figure 2.

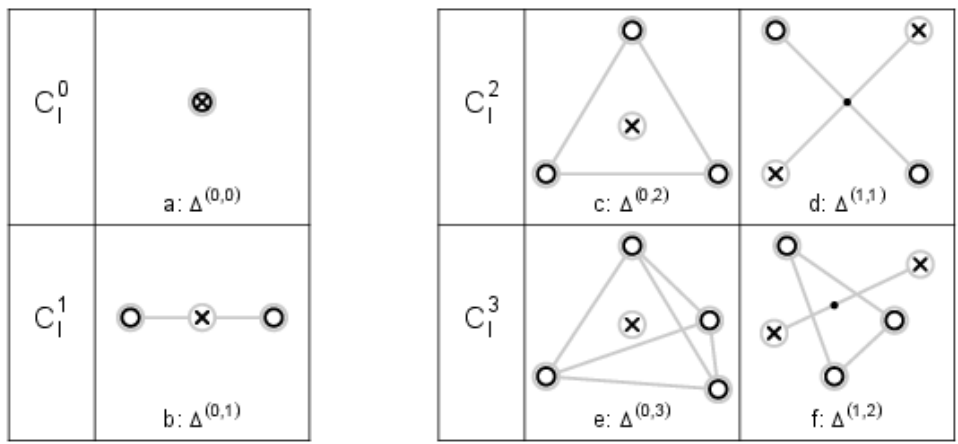

**Figure 2**: Two overlapping simplexes of dimension $d_1$ and $d_0$, $d = d_1 + d_0$

Configurations a and b appeared in Albert and Anderson (1984, Figure 1), but only as quasi-separation examples entailing 1D overlap configurations embedded in 2D.

The catalog depicted in Figure 2, and the others that follow, are limited to $0 \leq d_1 \leq d_0$ because the companion configurations, considered equivalents, are readily conjured by flipping responses. An account of all such configurations is made in the following:

**Theorem 3**. For each dimension $d = d_1 + d_0$, there are exactly $d+1$ unique Type I configurations.

**Proof**. For $d_1 = 0, 1, ..., d$, $d_0 = d - d_1$, $\Lambda^{(d_1, d_0)}$ (defined below) provides its matrix representation. The standard form for a Type I configuration, $\Lambda^{(d_1, d_0)}$, can be defined in terms of a unit n-simplex, $I^{(n)}$. Let $I_n$ be the identity matrix of order $n > 0$. This standard is described in Table 7.

**Table 7**. The standard Type I configuration, $\Lambda^{(d_1, d_0)}$, depicted as a matrix

| $d_1 = d_0 = 0$ | $d_1 > 0, d_0 = 0$ | $d_1 = 0, d_0 > 0$ | $d_1 > 0, d_0 > 0$ |
|---|---|---|---|
| $\begin{pmatrix} I^{(0)} \\ I^{(0)} \end{pmatrix}_{2 \times 1}$ | $\begin{pmatrix} I^{(d_1)} \\ 0_{1 \times d_1} \end{pmatrix}_{(d_1+2) \times d_1}$ | $\begin{pmatrix} 0_{1 \times d_0} \\ I^{(d_0)} \end{pmatrix}_{(d_0+2) \times d_0}$ | $\begin{pmatrix} I^{(d_1)} & 0_{(d_1+1) \times d_0} \\ 0_{(d_0+1) \times d_1} & I^{(d_0)} \end{pmatrix}_{(d+2) \times d}$ |

**Note**: $I^{(n)}$, the unit n-simplex, equals $(0)_{1 \times 1}$ or $\begin{pmatrix} I_n \\ -1_{1 \times n} \end{pmatrix}_{(n+1) \times n}$ for $n = 0$ or $n > 0$, respectively

**Theorem 4**. Type I configurations have a unique standard form $\Lambda^{(d_1, d_0)}$

**Proof**. Let $X = \begin{pmatrix} X^1_{n_1 \times d_1} \\ X^0_{n_0 \times d_0} \end{pmatrix}$ be a Type I configuration in matrix form, $D_{n \times n} = diag(u_1, u_0)$, $V_{n \times d} = D(X - 1_{(d+2) \times 1} S_{1 \times d})$ and $V_- = V_{-rows(n_1, n)}$. The interim form of $X$ is $V$, and the standard form of $X$ is $\Lambda_{n \times d} = VV_-^{-1}$.

The process of computing the standard form for the sample Type I configuration identified in Table 4 is illustrated in Table 8.

**Table 8**. Steps to put $W_1$, a Type I sub-configuration of $W_0$ (of Table 4) into Standard Form

| Data | $u_1^T$ | $Data_{y=1}$ | | | $u_0^T$ | $Data_{y=0}$ | | | $S^T = F^T \neq \varphi$ |
|---|---|---|---|---|---|---|---|---|---|
| Original | .75 | 11.0 | 4.6 | 8.1 | .4 | 10.4 | 4.1 | 3.5 | 10.0 |
| Form | .25 | 7.0 | 2.6 | 4.1 | .1 | 8.4 | 4.6 | 2.0 | 4.1 |
| X | | | | | .5 | 10.0 | 4.0 | 11.0 | 7.1 |
| | | | | | | | | | |
| Interim | 1/2 | .75 | .375 | .75 | 1/3 | .16 | .00 | -1.44 | 0 |
| Form | 1/2 | -.75 | -.375 | -.75 | 1/3 | -.16 | .05 | -.51 | 0 |
| V | | | | | 1/3 | .00 | -.05 | 1.95 | 0 |
| | | | | | | | | | |
| Standard | 1/2 | 1 | 0 | 0 | 1/3 | 0 | 1 | 0 | 0 |
| Form | 1/2 | -1 | 0 | 0 | 1/3 | 0 | 0 | 1 | 0 |
| $\Lambda$ | | | | | 1/3 | 0 | -1 | -1 | 0 |

**Note**: While a Type II configuration has no standard form, its interim form is usually simpler. The *intrm* function (Table 3) returns the interim form $V$.

Overlapping simplexes have the following property:

**Theorem 5**. The removal of one observation from a Type I configuration destroys overlap by creating either complete separation or no mixed results.

**Proof**. Let $X$ be a Type I configuration. First, assume $d_0 d_1 > 0$. In this case, it suffices to show complete separation results when the removed run occupies the last row of $X^1$. Let $\Lambda$ be the standard matrix form of $X$. Define $X_-$ and $\Lambda_-$ to be $X$ and $\Lambda$ without the last rows of $X^1$ and $\Lambda^{d_1}$, respectively. Also let $V$ be the interim matrix form of $X$ as defined in Table 7 and Theorem 4 and let $V_- = V_{-rows(n_1, n)}$. First, we prove the theorem for the standard form $\Lambda$. The last row of $\Lambda^{d_1}$ is $p_{1 \times d} = (-1_{1 \times d_1}, 0_{1 \times d_0})$. Let $q_{1 \times (n-1)} = p \Lambda_-^T$. The first $d_1$ entries of $q$ are minus one and the last $(d_0 + 1)$ entries are zero. Thus, $\Lambda_-^{d_1} p^T + .5 \prec 0$ and $\Lambda_-^{d_0} p^T + .5 \succ 0$, so the hyperplane $\{\Lambda_- | \Lambda_- p^T = -.5\}$ separates $\Lambda_-^{d_0}$ and $\Lambda_-^{d_1}$. Here, the strict curly inequality denotes component-wise inequality between two vectors. In terms of $X_-$, the hyperplane $\{X_- | X_- v^T = c\}$ separates $X_-^0$ and $X_-^1$, where $v = V_-^{-1} p$, $c = S \cdot v - .5 / \max(u_1)$ and $S = u_1 \cdot X_i^{d_1}$. If $d_1 d_0 = 0$, removal of the only Case or Non-Case observation from a Type I configuration makes $X^i = \phi$ for $i = 1$ and $i = 0$, respectively, which produces no mixed results.

**Corollary 2**. Type I configurations possess no lower dimensional overlapping sub-configurations.

For a Type I configuration, the interior of the intersection of its Case and Non-Case convex hulls contains one design point when $d_1 d_0$ is zero, and no design points when $d_1 d_0$ is non-zero.

**Theorem 6**. The removal of any one run from a Type I configuration does not reduce rank.

**Proof**. Table 8 indicates that every Type I configuration, in matrix form, has a standard form consisting of at least two rows. In each of the four possible configurations for $\Lambda^{(d_1, d_0)}$, the loss of one row does not reduce rank.

A regular n-simplex whose vertices are equidistant from each other and from the origin provides another standard form. This form, denoted by $E$ in Table 3, is returned by the function $equid(d_1, d_0)$. El-Gebeily and Fiagbedzi (2004) have an interesting discussion of the properties of the regular n-simplex. Boyd and Vandenberghe (2004, Chapter 2) provide an excellent introduction to convex sets, convex hulls, simplexes and separating hyperplanes.

Christmann and Rousseeuw (2001) explore the subject of measuring "regression depth", which they describe by two numbers:

$n_{overlap}$ is the smallest number of observations whose removal destroys overlap; and

$n_{complete}$ is the smallest number of observations whose removal yields complete separation.

For Type I configurations, it follows from Theorem 5 that $n_{overlap} = n_{complete} = 1$. For Type II configurations, $n_{overlap} = 1$ and $n_{complete}$ can be as large as $d + 1$ (when $n = 2(d + 1)$).

Numerical examples of Type I configurations of arbitrary dimension $d = d_1 + d_0$ can be readily obtained with the R function $stdf(d_1, d_0)$.

## 8. Type II Configurations ($d \leq 2$)

Two Type II configurations have been previously mentioned: multiple doubletons (Corollary 1) and runs 6-8 (Table 4, left) and runs 6-8 (Table 4, right). These examples indicate that, unlike Type I configurations, Type II configurations are hybrid compositions of two or more lower dimensional Type I and/or Type II configurations.

Based upon Table 6, a catalog of Type II configurations in two or fewer dimensions can be completed once the $n = 5$ case for $d = 2$ is worked out. Since the numbers are small, this is accomplished easily enough, e.g., by placing one X and four O's in the various cells of a tic-tac-toe board (and repeat the process with 2 X's and three O's; one $\otimes$, one X and three O's; and two $\otimes$'s and two O's). You wind up with the catalog of Type II configurations depicted in Figure 3 (with format annotations indicating how they appear to be composed).

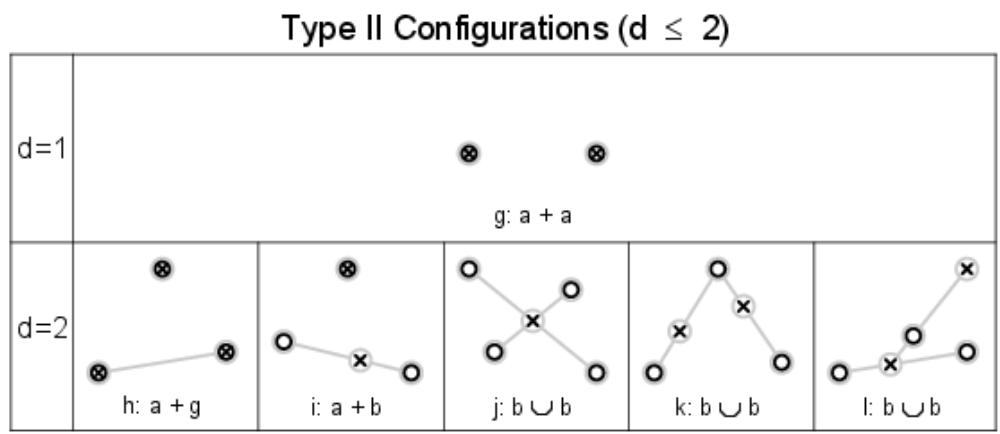

**Figure 3**. A catalog of Type II configurations (dimensions 1 and 2) generated by Add and Link

The two operators utilized in Figure 3 annotations, add ($+$) and link ($\cup$), are more fully described in the next three sections. They will be the two operations used to generate Type II configurations for the next dimension. Linked configurations will be seen to share one or more runs, whereas added configurations will not. Type II configurations constructed with one or more top-level add operations are therefore separable and linked Type II configurations are not. Note that the configurations being added may themselves be Type I configurations or Type II added or linked configurations.

A more robust annotation system is needed however, one that emphasizes dimensions rather than names. A notation is proposed, useful for creating add and link templates, beginning with the operator, followed by the dimensions involved (in increasing order). Exponents are employed to indicate repeated dimensions. For example, $+0^2 3$ indicates a Type II configuration generated by adding three configurations, two of dimension zero, and one of dimension three. This notation, though not as specific as $a + a + e$ or $a + a + f$ (configurations from Figure 2), conveniently puts them in the same tent, namely $+0^2 3$. When the operator is understood, $0^2 3$ is just as useful.

## 9. The Add Operation

The add operation is fashioned to assemble Type II configurations from lower dimensional Type I and/or Type II configurations. Pertinent to the operation, Leibnitz, Bernoulli and later Euler considered the problem of partitioning a natural number $n$ into a sum of two or more non-increasing natural numbers (Sierpinski, 1964, p. 400). Table 9 enumerates the unique partitions for the smallest natural numbers $n = d + 1$. The partitions are needed to iteratively populate catalogs containing Type II Add configurations.

**Table 9**. Add Template - A collection of formats to construct Type II Add Configurations in the form $C_{II}^d = \underset{i=1:m}{+} C_{I/II}^{d_i}$ where $d = \sum_{i=1:m} d_i + m - 1$

| $d$ | Additive Partitions of $d+1$ | | | | | Type II Add $(+)$ Formats Indicating Dimensions of the $m$ Configurations | | | | | | $g(d+1)$ |
|---|---|---|---|---|---|---|---|---|---|---|---|---|
| 1 | $1^2$ | | | | | $0^2$ | | | | | | 1 |
| 2 | 12 | $1^3$ | | | | 01 | $0^3$ | | | | | 2 |
| 3 | 13 | $2^2$ | $1^2 2$ | $1^4$ | | 02 | $1^2$ | $0^2 1$ | $0^4$ | | | 4 |
| 4 | 14 | 23 | $1^2 3$ | $12^2$ | $1^3 2$ | $1^5$ | 03 | 12 | $0^2 2$ | $01^2$ | $0^3 1$ | $0^5$ | 6 |
| 5 | 15 | 24 | $1^2 4$ | $3^2$ | 123 | | 04 | 13 | $0^2 3$ | $2^2$ | 012 | | |
| 5 | $1^3 3$ | $2^3$ | $1^2 2^2$ | $1^4 2$ | $1^6$ | | $0^3 2$ | $1^3$ | $0^2 1^2$ | $0^4 1$ | $0^6$ | | 10 |

**Note**: Enumerating the unique additive partitions and their number $g(d+1)$ was automated by Hankin (2006) in the R package *partitions*. The Type II Add formats are the respective partition components minus one.

Almost all of the lower dimensional formats involve addition of one or more 0-dimentional configurations. These additions are straightforward. Since it's less clear how to create examples when all sub-dimensions are non-zero, a sample implementation of the $1^3$ format (for $d = 5$) is included in the Appendix.

## 10. The Link Operation

The goal in this section is to characterize how 3D Type II configurations can be assembled by linking Type I configurations together. The first example considers linking three configuration $b$'s, depicted in Figure 2, together. All such links produce overlapping configurations, provided of course they are neither colinear nor coplanar. However, only four of them are minimal and hence Type II configurations. They are depicted in Figure 4.

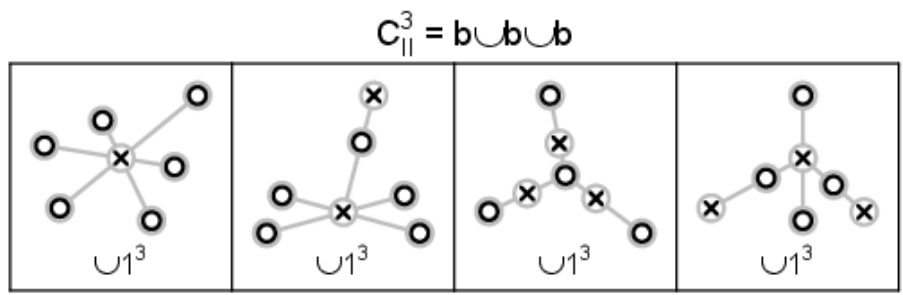

**Figure 4**. Four 3D Linked-Type II configurations found by linking three configuration $b$'s

The configurations that are not minimal are depicted in the two left panes of Figure 5. Each morph into added-Type II configurations, but only after the stricken run, depicted in Figure 5, is dropped. This indicates that certain exclusion rules are in play.

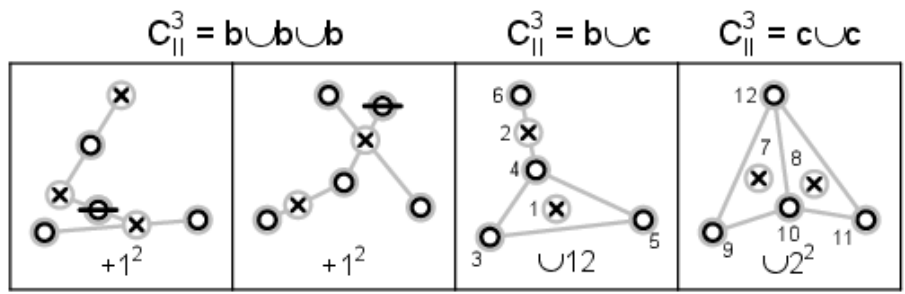

**Figure 5**. Four 3D Type II configurations found by linking various Type I configurations

The next two examples, depicted in the two right panes of Figure 5, are instances of linked-Type II configurations, first $b$ with $c$, then $c$ with $c$. Run numbers were added to facilitate further discussion of the exclusion rule.

Suppose you wanted to link another Type I configuration to the one depicted in pane three (or four) of Figure 5 to make a 4D Type II configuration. Runs 6 and 2 can't be linked to any other Type I configuration, but run 4 can. Run 4 can also be involved with another 2D link, e.g., rotate the 1-3-4-5 configuration about its 4-5 edge into 4D. This maneuver adds two runs, the 4D rotated versions of runs 3 and 1. You could not however rotate the 1-3-4-5 configuration about the 3-5 edge to get a linked-Type II configuration, due to the presence of the run 4 link with $b$. Similarly, you could rotate configuration 8-10-11-12 about its 10-12 edge, but not about its 10-11 or 11-12 edges, or any of its vertices. A configuration b can be linked successfully at runs 10 or 12, but not to any of the other four runs.

The exclusion rule is clear: each Type I configuration of dimension $d_i$ involved in a linked-Type II configuration has at most $d_i$ linked runs. Said another way, every Type I configuration involved in a linked Type II configuration maintains **_at least two unlinked runs_**.

A template of formats governing the construction of link Type II configurations is developed in Table 10. The formats depend on the maximum of the dimensions involved, $d_{max} = \max(d_i)$, $i = 1, 2, \ldots, m$, and the number of Type I components in the link, $m \geq 2$. These quantities, along with a scope index $s$, $0 \leq s \leq d_{max} - 1$, will determine the span of dimensions and sizes of the resulting configurations, according to the formulae: $d = d_{max} + m - 1 + s$ and $n = d + m + 1$. The link formats are described in Table 10.

**Table 10**. Link Template - A collection of formats, indexed by $s = 0, 1, \ldots, d_{max} - 1$, to construct Type II Link Configurations of the form $C_{II}^d = \underset{i=1:m}{\cup} C_I^{d_i}$

| $d_{max}$ | $m$ | $d$ | $(1:d_{max})_{m-1} \times d_{max}$ | Link Formats | | | | | | | |
|---|---|---|---|---|---|---|---|---|---|---|---|
| 1 | 2 | 2 | $(1:1)_1 \times 1$ | $1^2$ | | | | $s=0$ | | | |
|   | 3 | 3 | $(1:1)_2 \times 1$ | $1^3$ | | | | | | | |
|   | 4 | 4 | $(1:1)_3 \times 1$ | $1^4$ | | | | | | | |
|   | 5 | 5 | $(1:1)_4 \times 1$ | $1^5$ | | | | | | | |
|   | ⋮ | ⋮ | ⋮ | ⋮ | ⋮ | ⋮ | ⋮ | ⋮ | ⋮ | | |
| 2 | 2 | $3+s$ | $(1:2)_1 \times 2$ | 12 | $2^2$ | | | $s \leq 1$ | | | |
|   | 3 | $4+s$ | $(1:2)_2 \times 2$ | $1^2 2$ | $12^2$ | $2^3$ | | | | | |
|   | 4 | $5+s$ | $(1:2)_3 \times 2$ | $1^3 2$ | $1^2 2^2$ | $12^3$ | $2^4$ | | | | |
|   | ⋮ | ⋮ | ⋮ | ⋮ | ⋮ | ⋮ | ⋮ | ⋮ | | | |
| 3 | 2 | $4+s$ | $(1:3)_1 \times 3$ | 13 | 23 | $3^2$ | | $s \leq 2$ | | | |
|   | 3 | $5+s$ | $(1:3)_2 \times 3$ | $1^2 3$ | 123 | $2^2 3$ | $13^2$ | $23^2$ | $3^3$ | | |
|   | ⋮ | ⋮ | ⋮ | ⋮ | ⋮ | ⋮ | ⋮ | ⋮ | ⋮ | | |
| 4 | 2 | $5+s$ | $(1:4)_1 \times 4$ | 14 | 24 | 34 | $4^2$ | $s \leq 3$ | | | |
|   | ⋮ | ⋮ | ⋮ | ⋮ | ⋮ | ⋮ | ⋮ | ⋮ | | | |

**Note**: The notation $(1:n)_r$ represents all subsets of size $r$ that can be selected without regard to order from the set $\{1, 2, \ldots, n\}$, with replacement.

Table 10 was organized with telescoping blocks of formats indexed by a scoping number $s$. The scope $s$ indicates the span of dimensions and sizes for which the various link formats apply. The table was originally envisioned without scope. When the $2^2$ format was found capable of generating 4D configurations, it had to be accordingly reworked. Six $2^2$ scope-one 4D examples are discussed in Section 12 and depicted in Figure 9.

The 3D linked-Type II configurations depicted in Figure 4 and Figure 5 (two right panes) are just a few examples generated in accordance with $d = 3$ link formats.

## 11. Add and Link Overview

Specs for add and link Type II configurations are summarized in Table 11.

**Table 11**. Type II Configuration Specifications for the Add and Link Formats

| | Form | Constraints | | Dimension ($d$) | Size ($n$) |
|---|---|---|---|---|---|
| Add | $C_{II}^d = \underset{i=1:m}{+} C_{I/II}^{d_i}$ | $m > 1$ | | $\underset{i=1:m}{\sum} d_i + m - 1$ | $\underset{i=1:m}{\sum} n_i$ |
| Link | $C_{II}^d = \underset{i=1:m}{\cap} C_I^{d_i}$ | $m > 1$ | $\left\lvert C_I^j \cap C_I^k \right\rvert = min(j,k)$ | $d_{max} + m - 1 + s$ | $d + m + 1$ |

Figure 6 indicates how the add and link templates of Tables 10 and 11 apply to populating higher dimensional catalogs of Type II configurations.

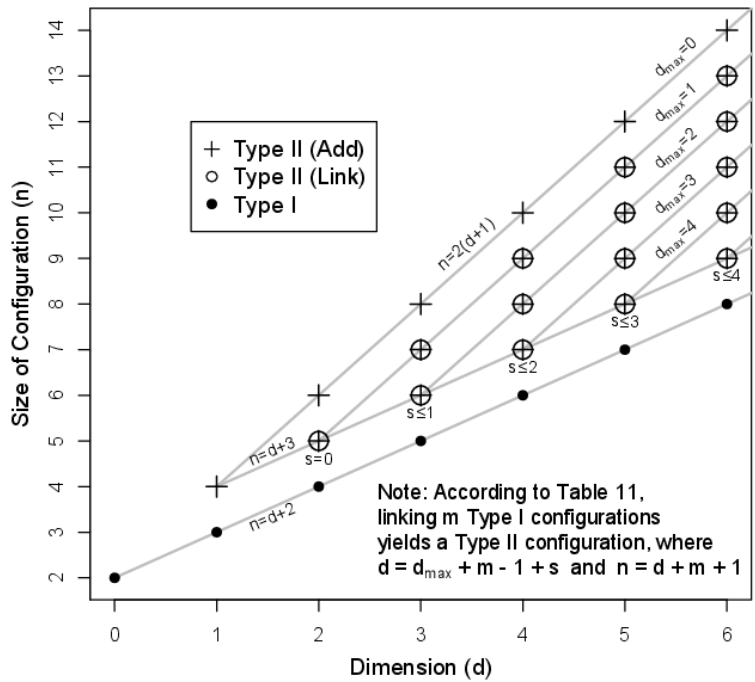

**Figure 6**. $n \times d$ Relationships, by Type and Format

## 12. Type II Add and Link Catalogs ($d = 3$)

To construct the add catalog for $d = 3$, we refer to Table 9 and go through its four formats. We conclude there's one way each to assemble $+0^4$ and $+0^2 1$; five ways to assemble $+02$; and two ways to assemble $+1^2$. The nine configurations are depicted in Figure 7.

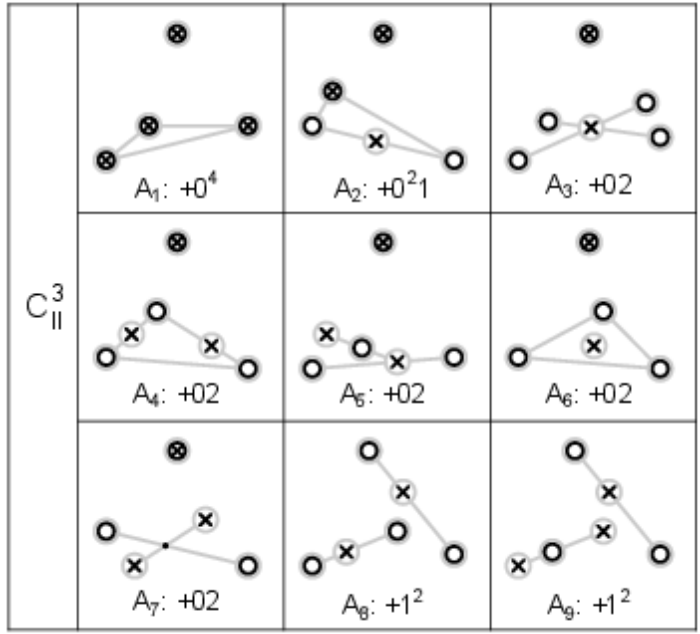

**Figure 7**. A catalog of 3D Added-Type II configurations obtained via the four Add formats

To construct the link catalog for $d = 3$, we refer to Table 10 and go through its three formats. We find there are six ways to assemble $\cup 12$; seven ways to assemble $\cup 2^2$; and four ways to assemble $\cup 1^3$. The seventeen configurations are depicted in Figure 8.

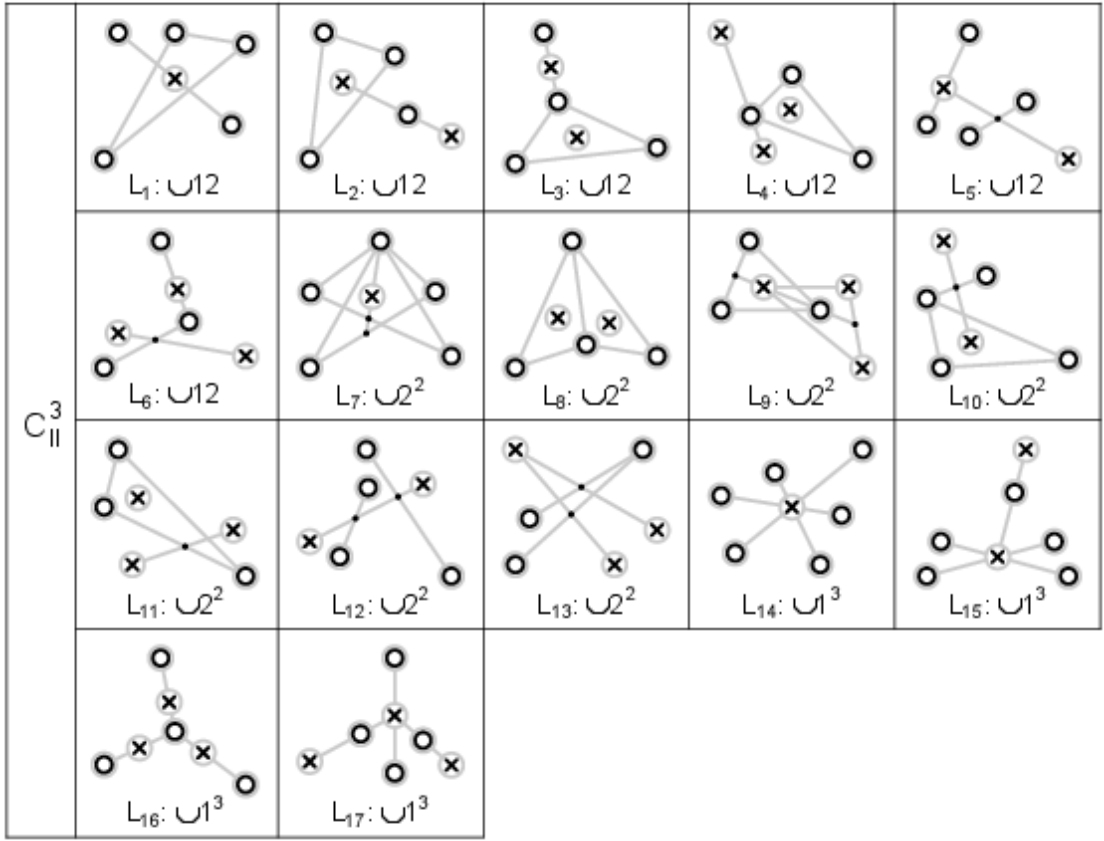

**Figure 8**. A catalog of 3D Linked-Type II configurations obtained via the three Link formats

    Close examination of Figures 3, 7 and 8 reveals that removing one run from separable Type II configurations, i.e., ones obtained via an add operation, creates quasi separation, whereas, for linked configurations possessing no singletons (or one singleton), the removal of one run produces quasi separation and complete separation (or no-mixed results), respectively. Additionally, all of the link formats required to assemble Figures 3 and 8 are scope-zero ( $s = 0$ ). The first scope-one format occurs in 4D, with the $2^2$ format specified in Table 10. Six configurations adhere to this format and are depicted in the left pane of Figure 9, where example sub-dimensions which support 4D are annotated next to each run.

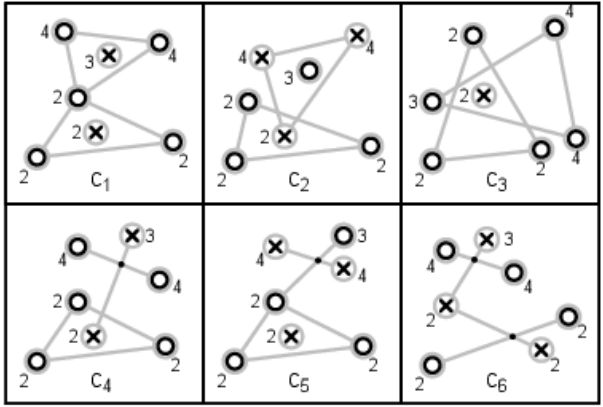

Configuration $c_6$ (left) is very similar to $c_7$ (below), a 4D $s=0$ configuration in $\cup 23$ format ($d \cup f$ of Figure 2)

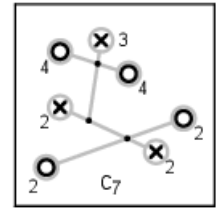

**Figure 9**. Six 4D Type II configurations in the first scope-one link format (left pane)

Notice there is only one 3D run in configuration $c_7$ (right pane of Figure 9). It happens to be the only run in the configuration that can't accommodate another link.

### 13. Completeness of the $d=3$ Catalog

The add portion of our 3D Type II catalog is depicted in Figure 7. Its completeness depends in a straightforward manner (described in Table 9) upon completeness of the lower dimensional catalogs depicted in Figure 2 (Type I) and Figure 3 (Type II). As such, add catalog completeness rests on establishing link catalog completeness. The 2D and 3D linked configurations depicted in Figures 3 and 8 contain three ($j$, $k$ and $l$) and seventeen ($L_i$, $i=1,\ldots,17$) configurations, respectively. In both cases, one may ask: are there no others?

Both questions are settled in the affirmative by conducting a systematic search over a lattice containing design points of a basis configuration of dimension $d$, plus two more design points of dimension $d+1$. Such a search suffices because the link specs described in Tables 8 and 9 imply adding two $(d+1)$-dimensional points to a $d$-dimensional Type I or to a Linked-Type II configuration is the way to populate the next link Type II catalog. The search procedure utilized in the 3D case is presently described in detail. The 2D search is similar, but much simpler.

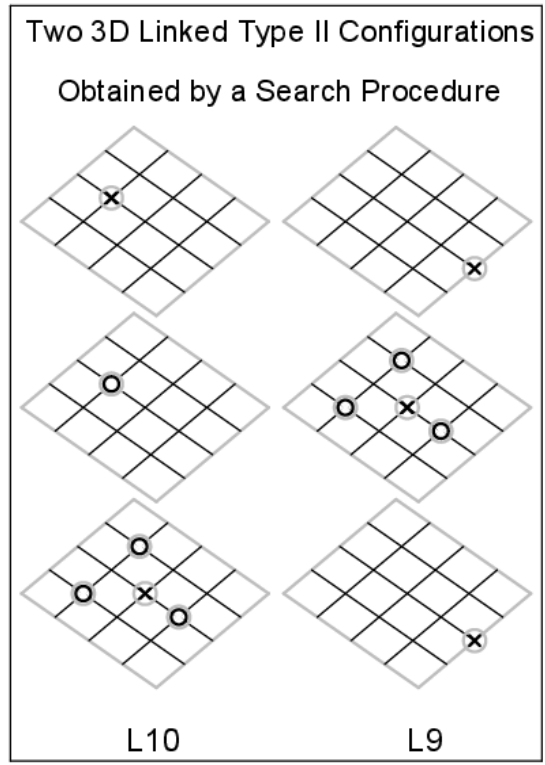

**Figure 10**. Addressing Completeness of Figure 8

In Figure 10, two search "hits" are depicted, L10 and L9 of Figure 8. In the L10 example, a 2D Type I, configuration *c*, is placed in the inner nine positions of the bottom lattice. The inner nine positions also accommodate the other 2D Type I configurations, namely *d*, as well as the linked-Type II configurations *j*, *k* and *l* of Figure 3. The middle lattice lies in the *xy* plane and is centered at the origin. The top and bottom lattices are parallel to the middle one, shifted up and down by one unit, respectively. One run is placed in each of the two empty lattices, in the same positions. Twenty-five such paired positions are tried, and for each try, four response pairs are assigned in turn: (0,0), (0,1), (1,0) and (1,1). In all, for each example, one hundred configurations are generated. An R script was written (*tictactoe*) to zip through each of them, first, with configuration *c* as the basis configuration, once located in the bottom plane, then in the middle plane. This process was repeated with configurations *d*, *j*, *k* and *l* serving, in turn, as the basis configuration. All configurations were tested for Type II-ness, and those meeting the criteria were further examined by the R function, *config3*, to see if they have already been cataloged or if they are new (i.e., need to be added).

The five by five lattice used in the search was deemed to be just large enough to catch any Type II linked-configurations that might otherwise escape detection. The need for a larger than three-by-three base-lattice was made clear in the L9 example of Figure 10, which indicates how the last Type II linked 3D configuration ($L_9$ of Figure 8) was actually discovered by the author.

The results of the systematic search conducted by *tictactoe* are summarized in Table 12.

**Table 12**. Completeness Verification of Figure 8, the 3D Type II Link Catalog

| Basis | Location | Overlap | Type II | New | ID | Repeats |
|---|---|---|---|---|---|---|
| c | Bottom | 50 | 14 | 0 | $L_2, L_3, L_8, L_{10}$ | 1, 3, 1, 9 |
| c | Middle | 50 | 14 | 0 | $L_1, L_4, L_7, L_9, L_{11}$ | 1, 3, 6, 3, 1 |
| d | Bottom | 50 | 14 | 0 | $L_6, L_{11}, L_{13}$ | 4, 2, 8 |
| d | Middle | 50 | 14 | 0 | $L_5, L_{10}, L_{12}$ | 4, 8, 2 |
| j | Bottom | 50 | 1 | 0 | $L_{17}$ | 1 |
| j | Middle | 50 | 1 | 0 | $L_{14}$ | 1 |
| k | Bottom | 50 | 1 | 0 | $L_{15}$ | 1 |
| k | Middle | 50 | 1 | 0 | $L_{16}$ | 1 |
| l | Bottom | 50 | 1 | 0 | $L_{16}$ | 1 |
| l | Middle | 50 | 1 | 0 | $L_{17}$ | 1 |

Since no new linked-Type II configurations were discovered in the systematic search, we may conclude that the $d < 4$ catalogs of minimal configurations presented in this work (Figures 2, 3, 7 and 8) are complete.

## 14. Quasi-Separation

The add and link templates given in Tables 9 and 10 indicate which sub-structure dimensions lead directly to Type II higher dimensional configurations. The individual sub-structures supporting Type II overlap in dimension $d$ are, by definition, quasi-separated configurations. Having referred to them as lower dimensional Type I and Type II configurations didn't sufficiently highlight one important fact: **Type II configurations are either added or linked quasi-separated configurations.**

    Quasi-completely separated data is said to be "less well-known to statisticians" (Candès and Sur 2020) and problematic because "iterative procedures for maximizing the likelihood tend to satisfy their convergence criterion before revealing any indication of separation" (Konis 2007). To follow up on this point Konis made, the *rquaz* function was used to generate quasi-separated data sets of arbitrarily large size by sandwiching a *d*-dimensional quasi-separated configuration between a randomly generated *d*-dimensional completely-separated data set. Large quasi-separated data sets like these were used to compare how *elstat* and *safetest* (featuring *safeBinaryRegression*) perform. The two R functions were found to work as advertised, provided compatible results, but did differ in the one respect: *safetest* is a pass/fail test for plain separation whereas *elstat* is more nuanced by differentiating between quasi-separation, complete separation and no-mixed results. Details about the comparisons are provided in the supplementary materials (*el4overlapSup.pdf*, Sections 23 and 24).

## 15. Concluding Remarks

In this note tools to generate and analyze sensitivity test data using the Neyer and 3pod sequential procedures as implemented in *gonogo* have been discussed. The need to develop similar procedures capable of dynamically varying two factors was highlighted as a state-of-the-art sensitivity testing design problem. Silvapulle's $S \cap F \neq 0$ overlap criterion was shown to be an empirical likelihood test that the mean displacement vector $\bar{\delta}$ is zero, rendering it testable via use of the *elstat* R-function. In the remainder of paper minimal configurations which account for overlap instances in dimensions fewer than four were explored. Type I and Type II (added and linked) configurations were distinguished, classified and visually cataloged. Rules to generate higher dimensional examples were also described.

Some unsuccessful attempts were made to extend the use of *emplik* to measure "regression depth" as discussed in Christmann and Rousseeuw (2001). This should have been expected since the problem is cited to be "essentially NP-hard" and requires a stochastic approach. While examining the findings of the authors, a few discrepancies were found. Under "important cases" pertaining to Finney's vasoconstriction data (Finney 1947), run triad 4, 18, 39 (not 4, 18, 29) is "important" and run triad 4, 18, 24 is not "important" (contrary to what was claimed). The minor errors occur in both Table 4 and Table 5, where "important" means removal of the cases destroys overlap and removal of the cases creates complete separation, respectively.

In a tutorial side-discussion of 1D overlap (Owen and Roediger, 2014), we stated that the four odd "corner cases" described in its equation (3), satisfied Silvapulle's $S \cap F \neq \phi$ overlap condition. They do not. Each are equivalent to a quasi-separated configuration that results from removing one run from the 1D Type II configuration g depicted in Figure 3. The *faux pas* had no bearing on any of the papers main results.

The empirical likelihood methodology for assessing overlap status should be extendable to multinomial responses along the same lines as originally suggested by Albert and Anderson (1984).

## 16. Rcode and documentation

The associated R suite of functions and documentation are posted with file names beginning with *el4Overlap* on Jeff Wu's personal website: https://www2.isye.gatech.edu/~jeffwu/sensitivitytesting/ .

### Appendix. Constructing Added Type II Configurations in Matrix Form

To construct add-examples using the template of Table 9, the matrix representations of the sub-configurations must be ensconced in a full-rank matrix having the correct number of columns, namely the tabled value of $d$. The trick is to make each sub-configuration rank deficient (i.e., higher dimensional) and remedying each deficiency by having other sub-configuration(s) occupy the extra dimensions (i.e., by sharing columns). The idea is made clearer as follows. Let $s$ be the number of shared rows in a matrix representation of an add format, say $i^n$ of dimension $d = n(i+1) - 1$ (by construction in Table 9). Also let $h$ be the number of extra dimensions added to each sub-configuration of dimension $i$. Another calculation of the total number of columns involved is: $d = i + s + (n-1)(i + h - s)$. Equating the two values gives $(n-1)(h-1) = (n-2)s$, which means $s = h = n-1$ since $n-1$ and $n-2$ are relatively prime.

The above details how add formats are used to create examples of Type II configurations in a column-wise-compressed matrix form. An example is provided in Table A1.

**Table A1**. Example Add Type II configuration in list form based on the $+1^3$ format

| *i* | *x* | | | | | *y* |
|---|---|---|---|---|---|---|
| 1 | 0 | 0 | 0 | | | 1 |
| 2 | 1 | 1 | 1 | | | 0 |
| 3 | -1 | -1 | -1 | | | 0 |
| 4 | | 2 | 3 | 1 | | 1 |
| 5 | | 1 | 5 | 2 | | 0 |
| 6 | | 4 | -1 | -1 | | 0 |
| 7 | | | 1 | 2 | 3 | 1 |
| 8 | | | 3 | 8 | -1 | 0 |
| 9 | | | 0 | -1 | 5 | 0 |

**Note**: The non-blank rows of $x$ are: $(0,0,0)+(1,1,1)k$ for $k=0,1,-1$ (rows 1-3); $(2,3,1)+(-1,2,1)k$ for $k=0,1,-2$ (rows 4-6); and $(1,2,3)+(1,3,-2)k$ for $k=0,2,-1$ (rows 7-9). Each $3\times 3$ sub-configuration is a 1D Type I configuration $b$ of Figure 2 embedded in 3D. The blank entries of $x$ are understood to be 0-filled.

**Acknowledgments**

The author would like to thank Art Owen for the many useful comments he provided, Jeff Wu for much encouragement, and Zhehui Chen for his technical support.